\documentclass[conference, a4paper]{IEEEtran}
\IEEEoverridecommandlockouts
\usepackage{cite}
\usepackage[english]{babel}
\usepackage{framed}
\usepackage[normalem]{ulem}
\usepackage{amsthm}
\usepackage{amssymb}
\usepackage[scr=dutchcal,cal=euler]{mathalfa}
\usepackage{enumerate}
\usepackage[utf8]{inputenc}
\usepackage{hyperref}
\hypersetup{
   colorlinks=true,
   linkcolor=blue,
   citecolor=blue,
   filecolor=magenta,      
   urlcolor=cyan,
}
\usepackage{amsmath,amssymb,amsfonts}
\usepackage{algorithm,algorithmic}
\usepackage{graphicx}
\usepackage{mathtools}
\usepackage{gensymb}
\usepackage{xcolor}
\usepackage{subcaption}
\usepackage{tabularx,booktabs}
\usepackage{multirow,multicol}
\usepackage{caption}
\usepackage{tikz}
\usetikzlibrary{arrows}
\def\BibTeX{{\rm B\kern-.05em{\sc i\kern-.025em b}\kern-.08em
    T\kern-.1667em\lower.7ex\hbox{E}\kern-.125emX}}

\tikzstyle{int}=[draw, fill=blue!20, minimum size=3em]
\tikzstyle{sum}=[draw, circle, minimum size=0.5em]
\tikzstyle{init} = [pin edge={to-,thin,black}]

\begin{document}

\title{System Identification on the Families of Auto-Regressive with Least-Square-Batch Algorithm\\
}

\author{Moh Kamalul Wafi$^{1}$ 
\thanks{$^{1}$Moh Kamalul Wafi is with Laboratory of Embedded and Cyber-Physical Systems, Department of Engineering Physics,
        Institut Teknologi Sepuluh Nopember, 60111, Indonesia,
        {\tt\small kamalul.wafi at its.ac.id}}%
}

\maketitle

\begin{abstract}
The theories of system identification have been highly elaborated so as to achieve the true system. This paper much discuses regarding the stochastic processes along with the divergent of whether or not the system has zero-mean under scenario of either white and coloured noise. The mathematical foundations, including mean, variance, covariance, optimal parameters, along with some modified scenarios among them, are presented in detail from various system along with some basic idea behind them. The families of auto-regressive (1) and (2) are compared both mathematical and simulation in order to obtain the best design approaching the true system, Moreover, the least-square algorithm is used to examine the effectiveness of some number of iteration along with "batch" theorem.
\end{abstract}

\begin{IEEEkeywords}
System Identification, Auto-Regressive, Least-Square Algorithm, Batch Theorem
\end{IEEEkeywords}

\section{Introduction}
System identification, an error-correcting training scheme, is becoming more important dealing with either a system or process requiring to estimate the states with certain unknown variables. Further research even has modified from classic statistical method to regularisation considering the unhealthy matrix of minimization scheme \cite{R1}, \cite{R2}. The idea of auto-regressive involves current and preceding data to be estimated with mutually independent per-sequence \cite{R3}. This model is then modified so as to obtain the best prediction, reducing the error. This is well-conducted by \cite{R4} proposing unknown number of epoch formula each sequence inducing the Markov process. Thanks to \cite{R5}, the very early idea in predicting one-step ahead using either the median and the log-normal process of distribution. Moreover, the procedural modification applying Lasso scenario is also a choice as presented in \cite{R6}. This research is to examine basic properties from the family of auto-regressive along with the benefit of some iteration $N$ and batch $\kappa$. Beyond that, the assumptions of zero-mean and its counterpart along with either white and coloured noise are also considered.

The first and foremost is the stability properties of the system. The Lyapunov scenario as the most famed algorithm is proposed in this project \cite{R7}. While there has been a broad modification in determining the stability of the system, including strong passivity in nonlinear system as stated in \cite{R8}, the mathematical foundation should be well-defined \cite{R9}. These basic information are then ready followed by some recursive or forecasting scheme with other bunch of methods. Least-square as prediction is to portray the best linear to show the data along with forecast the following data with respect to some instant time $t$, consisting the trend and the level of development. The more generalized with coloured noise system is presented in \cite{R10} referring to the original \cite{R11}. This paper comprises the problem formulation followed by the basic mathematical concepts. The next is the analytical result with three divergent scenarios ended by the numerical design and conclusion.

\section{Problem Formulation}
This paper presents the objectives of system identification:
\begin{enumerate}
\item To elaborate the functions of expected value, variance, and covariance in a stationary process via static representation
\item To describe mathematically the identification abilities of AR models of increasing order
\item To compare the theoretical results with its simulation with empirical Least Square (LS) estimation
\item To highlight the impact of coloured noises on the LS estimation
\end{enumerate}

\section{Basic Mathematical Theories}
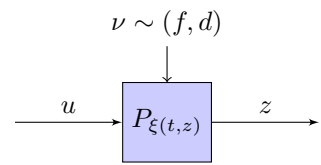
\begin{figure}[h]
\centering
\begin{tikzpicture}[auto,node distance=2.5cm,>=latex']
    \node [int, pin={[init]above:$\nu \sim (f, d)$}] (a) {$P_{\xi(t,z)}$};
    \node (b) [left of=a,node distance=2cm, coordinate] {a};
    \node [coordinate] (end) [right of=a, node distance=2cm]{};
    \path[->] (b) edge node {$u$} (a);
    \draw[->] (a) edge node {$z$} (end) ;
\end{tikzpicture}
\caption{The general dynamic systems}
\label{Fig1}
\end{figure}

General design of the system model is presented in Fig. (\ref{Fig1}) defining that $P$ is the system plant along with $\xi$ as its parameter. $\nu$ comprises either fault $(f)$ or disturbances $(d)$ whilst $u$ and $z$ denote as the accessible input and the controlled variable in turn. The space of possible results is denoted by $\mathcal{S}$ and its single result of $s$ is $\in \mathcal{S}$ while the events of certain interest are defined as $s_\epsilon$ which is also the subset of $\mathcal{S}$. Furthermore, the patterns of the output depend on the parameters on algebraic equation of the system $P_\xi$ resulting its roots either real or complex. The common features of the dynamics constitute that the $(t)$ continuous or $(z)$ discrete-mathematical structure is built by differential equation and it has been a-priori knowledge being found from the physical properties of the system. 
The applied random variable is a variable of $\mathcal{F}(s)$ obtaining values according to the result $s \in \mathcal{S}$ of a random trial through a function $\Theta(\bullet)$. The probability distribution function supplies the data on the random variable so that the equation of $F(x) = \mathcal{P}(v \leq x)$ can be denoted as,\newpage
\begin{equation}
	\mathcal{P}(v \in [a, b]) = F(b) - F(a)
\end{equation}
meaning that $\mathcal{P}$ is the specific probability distribution being made of function $F$ with points between $a$ and $b$ while the probability density of $f^\dagger$ creating an area of $\mathcal{P}(v \in [a, b])$ is defined as follows,
\begin{equation}
	f^{\dagger}(x) = \frac{\partial F}{\partial x}
\end{equation}
Furthermore, the expected value $\textbf{E}$ and its variance $\mathcal{Q}$ of a fault and disturbance are presented as,
\begin{align}
	\textbf{E}(v) &= \int_{-\infty}^{+\infty} x f^\dagger(x) \, dx\\
    \mathcal{Q}_v &= \int_{-\infty}^{+\infty} \left[x - \textbf{E}(v) \right]^2 f^\dagger(x) \, dx 
\end{align}
with standard deviation of $\sigma(v) = \sqrt{\mathcal{Q}_v}$. Keep in mind that, given two variables of $v_1(s)$ and $v_2(s)$, the results of $v(s)$, $\textbf{E}(v)$, and $\mathcal{Q}_v$ are $v(s) = v_1(s) + v_2(s)$, $\textbf{E}(v) = \textbf{E}(v_1) + \textbf{E}(v_2)$, and
\begin{equation}
	\mathcal{Q}(v) \neq \mathcal{Q}(v_1) + \mathcal{Q}(v_2)
\end{equation}
in turn, with $\mathcal{Q}_v = \textbf{E}\left\lbrace [v - \textbf{E}(v)][v - \textbf{E}(v)]^T \right\rbrace$. A Gaussian random variable exists if the average $\mu = \textbf{E}(v)$ and variance $\mathcal{Q}$ or $\sigma^2$ satisfies:
\begin{equation}
	g(x) = \frac{1}{\sqrt{2 \pi} \sigma} e^{-k} \longrightarrow k = \frac{(x - \mu)^2}{2\sigma^2}
\end{equation}
with respect to the properties of the $v$ in terms of correlation and independent, again, the two random variables $v_1$ and $v_2$ are \textbf{uncorrelated} if
\begin{equation}
	\textbf{E}\left\lbrace [v_1 - \textbf{E}(v_1)][v_2 - \textbf{E}(v_2)]^\top \right\rbrace = 0
\end{equation}
whereas they are called as independent if
\begin{equation}
	\mathcal{P}(v_1, v_2) = \mathcal{P}(v_1) \mathcal{P}(v_2)
\end{equation}
hence, the stochastic process originally requires the portrayal of probability distribution and is a collection of infinite random variable sequenced with respect to time. A stochastic process of $v(t)$ is called \textbf{white} being defined as $v \sim \mathcal{W}\mathcal{N}(0, \lambda^2)$ if $\textbf{E}[v(t)] = 0$ and
\begin{equation}
	\mathcal{Q}(x) = \begin{cases}
	\lambda^2, & x = 0\\
    0, & x \neq 0
	\end{cases}
\end{equation}
Beyond that, apart from examine the unbiased estimation, the minimization of the variance $\Phi$ could be tested. The estimator of $\bar{\theta}^{1}$ is better than that of $\bar{\theta}^{2}$ if $\Phi_{\bar{\theta}^1} \leq \Phi_{\bar{\theta}^2}$ or, if $\bar{\theta}^i$ is a vector, then $\Phi_{\bar{\theta}^2} - \Phi_{\bar{\theta}^1} \geq 0$. The asymptotic characteristic is affected by the number of simulated data $N$ which also influence the better estimation and reduce the uncertainty. Suppose three different variances $\Phi_m$ with the same expected value \textbf{E} such that $\textbf{E}[\bar{\theta}^i] = \theta$ with $i = 1\to m$ and $m_3 > m_2 > m_1$, the estimate $\bar{\theta}^m$ is said good at converging to $\theta$ if 
\begin{align*}
    \lim_{m\to\infty} \Phi_{\bar{\theta}^m} = 0 \longrightarrow \lim_{m\to\infty} \textbf{E}\left[\lVert\bar{\theta}^m - \theta\rVert^2\right] = 0
\end{align*}
where $\bar{\theta}^m, \theta$ and $\lVert\bar{\theta}^m - \theta\rVert$ are the random and constant vector along with a scalar random with good mean variable in turn.

\section{Analytical Results}
This section acts as the numerical findings which are then compared to the simulations with Least Square algorithm being elaborated in the following section. There are three different scenarios. First is the basic with zero-mean and the second is to alter from zero-mean to non-zero mean yet the other conditions are kept the same and the following is to change the covariance instead of the mean.

\subsection{Independent and Zero-mean}
The dynamics of a stationary stochastic process $y(\bullet)$ is presented in the following Fig. (\ref{Fig2}) in which $q(\bullet)$ and $v(\bullet)$ are independent, such that:
\begin{figure}[h!]
\centering
\begin{tikzpicture}[node distance=2.5cm,auto,>=latex']
    \node [int] (a) {$\dfrac{1}{1 - \lambda z^{-1}}$};
    \node (b) [left of=a,node distance=2cm, coordinate] {a};
    \node [sum, pin={[init]above:$v \sim (f, d)$}] (c) [right of=a] {$+$};
    \node [coordinate] (end) [right of=c, node distance=2cm]{};
    \path[->] (b) edge node {$q(t)$} (a);
    \path[->] (a) edge node {} (c);
    \draw[->] (c) edge node {$y(t)$} (end) ;
\end{tikzpicture}
\caption{Dynamic systems of the proposed model}
\label{Fig2}
\end{figure}
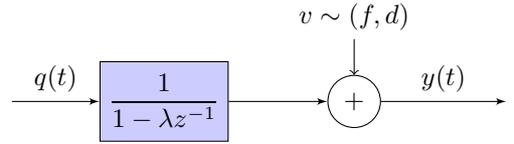\\
where $q \sim \mathcal{W}\mathcal{N}\left(\bar{q}, \delta^2\right)$, $v \sim \mathcal{W}\mathcal{N}\left(\bar{v}, \xi^2\right)$, with $\left[\bar{q},\bar{v}\right] = 0$ and $\lambda \in \mathbb{R}, |\lambda| < 1$. It can clearly be shown from the Fig. (\ref{Fig2}) that there have been two variables which influence the point of output $y(t)$ so that, by applying some algebra shown in \textbf{Appendix A} from inversing the $z-$transform to (\textit{time}) $t-$domain, it follows that:
\begin{align}
    y(t) = \frac{1}{1 - \lambda z^{-1}}q(t) + v(t)
\end{align}
can be then written as the true system of the following formula, therefore
\begin{equation}
	\mathcal{S}: y(t) = \lambda y(t - 1) + q(t) + v(t) - \lambda v(t - 1)
\end{equation}
As the inputs process of $q(\bullet)$ and $v(\bullet)$ are stationary and the point of $\lambda$ lies strictly inside the unit circle as well as $|\lambda| < 0$ which makes the value of $|z| < 0$, the dynamic model generating the stochastic process is asymptotically stable which implies that the steady state stochastic process $y(t)$ is stationary. Furthermore, since the zero-mean, the expected value $\textbf{E}[y(t)] = \bar{y}$ computed in \textbf{Appendix B} is zero,
\begin{align}
    \textbf{E}[y(t)] = \bar{y} = 0 \longrightarrow \left[\bar{q},\bar{v}\right] = 0
\end{align}
or substituting the two mean of the independent $[q,v]$ by the end of the solution, the formula is then finished as $\bar{y} = \lambda\bar{y}$, concluding that $\bar{y} \rightarrow 0$ and $\lambda\neq 0$. As for the variance of $y(t)$ denoted as $\Phi_y$, it is solved in the \textbf{Appendix C}, hence
\begin{align}
    \Phi_y =\; & \frac{\delta^2 + (1 - \lambda^2) \xi^2}{1 - \lambda^2}
\end{align}
Since the expected value $(\bar{y})$ is zero, the covariance function $\Psi_y$ with $\tau = 0, 1, \ldots$ is exactly the same as the correlation function being calculated in \textbf{Appendix D} and \textbf{D$^\star$}, with
\begin{align}
    \Psi_y(\tau) =\; & \textbf{E}\left\lbrace\left[y(t) - \bar{y}\right]\left[y(t - \tau) - \bar{y}\right]\right\rbrace 
\end{align}
such that,
\begin{align}
    \Psi_y(\tau) = \begin{cases}
    \dfrac{\delta^2}{1 - \lambda^2} + \xi^2 &, \tau = 0\\
    \\
    \dfrac{\delta^2}{1 - \lambda^2}\lambda^{\tau} &, \tau > 0
    \end{cases}
\end{align}
Furthermore, the analysis of the proposed dynamical process $y(\bullet)$ is required using a PEM identification algorithm of the family AR(1), such that
\begin{align}
    \mathbscr{F}_1(\theta_1): y(t) = \phi_1 y(t-1) + \psi(t)
\end{align}
and the corresponding stage of the design $\mathbscr{F}_1$ in estimation form is,
\begin{align}
    \hat{\mathbscr{F}}_1(\theta_1): \hat{y}(t|t-1) = \phi_1 y(t-1)
\end{align}
As for the higher-order of the family AR(2), the $\mathbscr{F}_2$ is denoted as follows,
\begin{align}
    \mathbscr{F}_2(\theta_2): y(t) = \phi_1 y(t-1) + \phi_2 y(t-2) + \psi(t)
\end{align}
along with its prediction form, such that
\begin{align}
    \hat{\mathbscr{F}}_2(\theta_2): \hat{y}(t|t-1) = \phi_1 y(t-1) + \phi_2 y(t-2)
\end{align}
where $\theta_1$ and $\theta_2$ equal to $\phi_1$ and $[\phi_1, \phi_2]^{\top}$ in turn along with suitable white noise of $\psi(t)$. Keep in mind that to compute the optimal values of those two $\left(\theta^{\star}_1, \theta^{\star}_2\right)$ with respect to $\left(\theta_1, \theta_2\right)$ respectively is to use the PEM algorithm and according to the asymptotic theory, the convergence to one of the minima of the two functions $\Gamma(\theta_1)$ and $\Gamma(\theta_2)$ can be guaranteed, such that
\begin{align}
    \Gamma(\theta_1) = \textbf{E}\left\lbrace\left[\varepsilon_{\theta_1}(t) \right]^2\right\rbrace =\; & \textbf{E}\left\lbrace\left[y(t) - \hat{y}(t|t-1)\right]^2\right\rbrace\\
    \Gamma(\theta_2) = \textbf{E}\left\lbrace\left[\varepsilon_{\theta_2}(t) \right]^2\right\rbrace =\; & \textbf{E}\left\lbrace\left[y(t) - \hat{y}(t|t-1)\right]^2\right\rbrace
\end{align}
where $\varepsilon_{\theta_1}(t)$ is the prediction error with suitable $y(t)$ as shown in the $\mathbscr{F}_1$ and $\mathbscr{F}_2$ and regarding the cost (optimal) functions of the two, they are defined as follows,
\begin{align}
    \frac{\partial\Gamma(\theta_1)}{\partial\phi_1} =\; 0 \quad \textrm{ and } \quad \frac{\partial\Gamma(\theta_2)}{\partial\phi_1\partial\phi_2} =\; 0
\end{align}
Furthermore, applying the concepts of [] above results in the optimality and recalling the $\Psi_y(\tau)$, the values of $\left(\theta^{\star}_1, \theta^{\star}_2\right)$ are written in the following according to the calculation being done in \textbf{Appendix E} and \textbf{F},
\begin{align}
    \theta^{\star}_1 \coloneqq\; & \phi_1 = \frac{\lambda\delta^2}{\delta^2 + (1-\lambda^2)\xi^2}\\
    \theta^{\star}_2 \coloneqq\; & \left[\phi_1, \phi_2\right]^{\top} = \begin{cases}
    \phi_1 = \dfrac{\lambda\delta^2\left(\delta^2 + \xi^2\right)}{\left(\delta^2 + \xi^2\right)^2 - \lambda^2\xi^4} \\
    \\
    \phi_2 = \dfrac{\lambda^2\delta^2\xi^2}{\left(\delta^2 + \xi^2\right)^2 - \lambda^2\xi^4}\end{cases}
\end{align}
Those optimum values then are implemented in the models of AR(1) and AR(2) as the best design of either families, $\mathbscr{F}_1(\theta^{\star}_1)$ with $\theta^{\star}_1\coloneqq\phi_1$ and $\mathbscr{F}_2(\theta^{\star}_2)$ with $\theta^{\star}_2\coloneqq\left[\phi_1, \phi_2\right]^{\top}$, approximating the true system. The variance of the prediction errors related to the two designs can be calculated from [] due to the fact that the expected values of the error from either families are zero, such that,
\begin{align}
    \bar{\varepsilon}_{\theta_1} =\; & \textbf{E}\left[y(t)\right] - \phi_1\textbf{E}\left[y(t-1)\right] = 0\\
    \bar{\varepsilon}_{\theta_2} =\; & \textbf{E}\left[y(t)\right] - \phi_1\textbf{E}\left[y(t-1)\right] - \phi_2\textbf{E}\left[y(t-2)\right] = 0
\end{align}
and the variance of the families are enlightened in \textbf{Appendix G} by substituting the covariance, therefore
\begin{align}
    \Phi_{\varepsilon_{\theta_1^\star}} =\; & \frac{\left(\delta^2 + \xi^2\right)^2 - \lambda^2\xi^4}{\delta^2 + \left(1-\lambda^2\right)\xi^2}\\
    \Phi_{\varepsilon_{\theta_2^\star}} =\; & \frac{\left(\delta^2 + \xi^2\right)^3 - \lambda^2 \left(\delta^2\xi + \xi^3\right)^2 + 2\lambda\delta^4\xi^2}{1 - \lambda^2}
\end{align}

\subsection{Independent and Non-Zero Mean}
Recalling the same dynamical process as Fig. (\ref{Fig2}), the $q$ and $v$ are no longer zero-mean, constituting $q \sim \mathcal{W}\mathcal{N}\left(\bar{q}, \delta^2\right)$ and $v \sim \mathcal{W}\mathcal{N}\left(\bar{v}, \xi^2\right)$ with $\bar{q}$ and $\bar{v}$ equal to 1 and 4 respectively. Keep in mind that when $q(\bullet)$ and $v(\bullet)$ are zero-mean, the process $y(\bullet)$ is also zero-mean and by contrast, if the two are not, this makes the expected value of the process is not zero anymore, recalling \textbf{Appendix B}, the $\bar{y}$ is,
\begin{align}
    \bar{y}\coloneqq\; \textbf{E}[y(t)] =\; & \lambda\bar{y} + \bar{q} + \bar{v} - \lambda\bar{v}\nonumber \\
    =\; & \bar{v} + \dfrac{\bar{q}}{1 - \lambda} \longrightarrow \frac{5-4\lambda}{1-\lambda}
\end{align}
and this also leads to the introduction of the new process $\check{y}(t)$, where $\check{y}(t) = y(t) - \bar{y}$, therefore
\begin{align}
    y(t) = \check{y}(t) + \bar{y}
\end{align}
such that,
\begin{align}
    \Phi_y = \textbf{E}\left\lbrace [y(t) - \bar{y}]^2 \right\rbrace = \textbf{E}\left\lbrace [\check{y}(t)]^2 \right\rbrace
\end{align}
However, the not zero-mean process of $y(t)$ does not yield the changes of the variance because the mean is subtracted from it. Likewise, due to the coincidence between the correlation function in the zero mean process $\check{y}(t)$ and the covariance of the original process, this new process does not give divergent results in terms of covariance, meaning that both parameters are on par with the originals, therefore
\begin{align}
    \Psi_y(\tau) =\; & \textbf{E}\left\lbrace\left[y(t) - \bar{y}\right]\left[y(t - \tau) - \bar{y}\right]\right\rbrace \nonumber\\
    =\; & \textbf{E}\left\lbrace\left[\check{y}(t)\right]\left[\check{y}(t-\tau) \right]\right\rbrace
\end{align}
Furthermore, the AR families analysis of $\mathbscr{F}_1$ and $\mathbscr{F}_2$ regarding the optimality $\left(\theta^{\star}_1, \theta^{\star}_2\right)$ which are affected by the covariance as mentioned in Eqs. (21) and (22) result in the similarities of the original processes with slight differences. Recalling the new process of $\check{y}(t)$, the cost function $\Gamma(\theta_1)$, $\Gamma(\theta_2)$ along with their estimates,
\begin{align}
    \hat{y}(t|t-1)_{\hat{\mathbscr{F}}_1} =\; & \phi_1 y(t-1) \nonumber\\
    =\; & \phi_1 \left[\check{y}(t-1) + \bar{y}\right]\\
    \hat{y}(t|t-1)_{\hat{\mathbscr{F}}_2} =\; & \phi_1 y(t-1) + \phi_2 y(t-2) \nonumber \\
    =\; & \phi_1 \check{y}(t-1) + \phi_1 \check{y}(t-2) + \left(\phi_1 + \phi_2\right) \bar{y}
\end{align}
respectively. The optimal value of $\left(\theta^{\star}_1, \theta^{\star}_2\right)$ is then obtained in \textbf{Appendix H} and \textbf{I}, such that
\begin{align}
    \theta^{\star}_1 \coloneqq\; & \phi_1 = \frac{\lambda\delta^2 + \left(1-\lambda^2\right) \bar{y}^2}{\delta^2 + \left(1-\lambda^2\right)\left(\xi^2 + \bar{y}^2\right)}
\end{align}
and $\theta^{\star}_2 \coloneqq\; \left[\phi_1, \phi_2\right]^{\top}$, with
\begin{align}
    \begin{bmatrix}\dfrac{\left[\lambda\delta^2 + \left(1-\lambda^2\right)\bar{y}^2 \right]\left(\delta^2 + \xi^2\right)}{\left(\delta^2 + \xi^2 + 2\bar{y}^2\right)\left(\delta^2 + \xi^2\right) - \xi^2\lambda^2\left(\xi^2 + 2\bar{y}^2\right) - 2\lambda\delta^2\bar{y}^2}\\[15pt]
    \dfrac{\lambda^2\delta^2\xi^2 + \left[\delta^2\left(1-\lambda\right)^2 + \xi^2\left(1-\lambda^2\right) \right]\bar{y}^2}{\left(\delta^2 + \xi^2 + 2\bar{y}^2\right)\left(\delta^2 + \xi^2\right) - \xi^2\lambda^2\left(\xi^2 + 2\bar{y}^2\right) - 2\lambda\delta^2\bar{y}^2}\end{bmatrix}
\end{align}
Likewise, these result are applied as the most optimum designs in either families approaching the true systems, while the variance of them are modified using these following mean of prediction error, such that 
\begin{align}
    \bar{\varepsilon}_{\theta_1} =\; & (1 - \phi_1)\bar{y}\\
    \bar{\varepsilon}_{\theta_2} =\; & (1 - \phi_1 - \phi_2)\bar{y}
\end{align}
and the results are showed in \textbf{Appendix J} with small divergent compared to the cost functions of $\left(\Gamma(\theta_1), \Gamma(\theta_2)\right)$ by the square of the expected values. Finally, under the initial assumption of the characteristics $q(\bullet)$ and $v(\bullet)$, these influence almost the whole systems leading to the shifted distribution.

\subsection{Assumption to Parameters of $\lambda, \delta^2, \xi^2$}
Turning to another design with similar dynamical process as Fig. (\ref{Fig2}), the $q$ and $v$ are set as zero-mean with certain true parameters, comprising $q \sim \mathcal{W}\mathcal{N}\left(0, 4\right)$ and $v \sim \mathcal{W}\mathcal{N}\left(0, 9\right)$. This design will not change the expected value with $\bar{y} = 0$ and the result makes the variance and covariance are counted the same as the originals (see \textbf{Appendix C} and \textbf{D}) with,
\begin{align}
    \Phi_y = 13.5
\end{align}
and,
\begin{align}
    \Psi_y(\tau) = \begin{cases}
    13.5 &, \tau = 0\\
    \\
    4.5\left(\dfrac{1}{3}\right)^{\tau} &, \tau > 0
    \end{cases}
\end{align}
Furthermore, since the model of AR(1) is affected by $\Psi_y(\tau)$, the optimal value of $\theta_1^{\star}$ and the best design of $\mathbscr{F}_1$ are
\begin{align}
    \theta_1^{\star}\coloneqq\phi_1 = \frac{1}{9} \longrightarrow \mathbscr{F}_1 : y(t) = \frac{1}{9}y(t-1) + \phi(t)
\end{align}
which is approaching the true system with variance $\Phi_{\theta_1^{\star}}$ equals to $13.333$. Likewise, as for the counterpart family of AR(2), the optimal value of $\theta^{\star}_2$ is
\begin{align}
    \theta^{\star}_2\coloneqq\; & \left[\phi_1, \phi_2\right]^{\top} = \left[\frac{39}{360}, \frac{1}{40}\right]^{\top}
\end{align}
so that the best model of AR(2) can be written as follows
\begin{align}
    \mathbscr{F}_2 : y(t) = \frac{39}{360}y(t-1) + \frac{1}{40}y(t-2) + \phi(t)
\end{align}
with the variance $\Phi_{\theta_2^{\star}}$ equals to $13.325$. From the model, the AR(2) is slightly better in terms of small variance with $0.008$ difference compared to its counterpart. 

\begin{figure*}[t!]
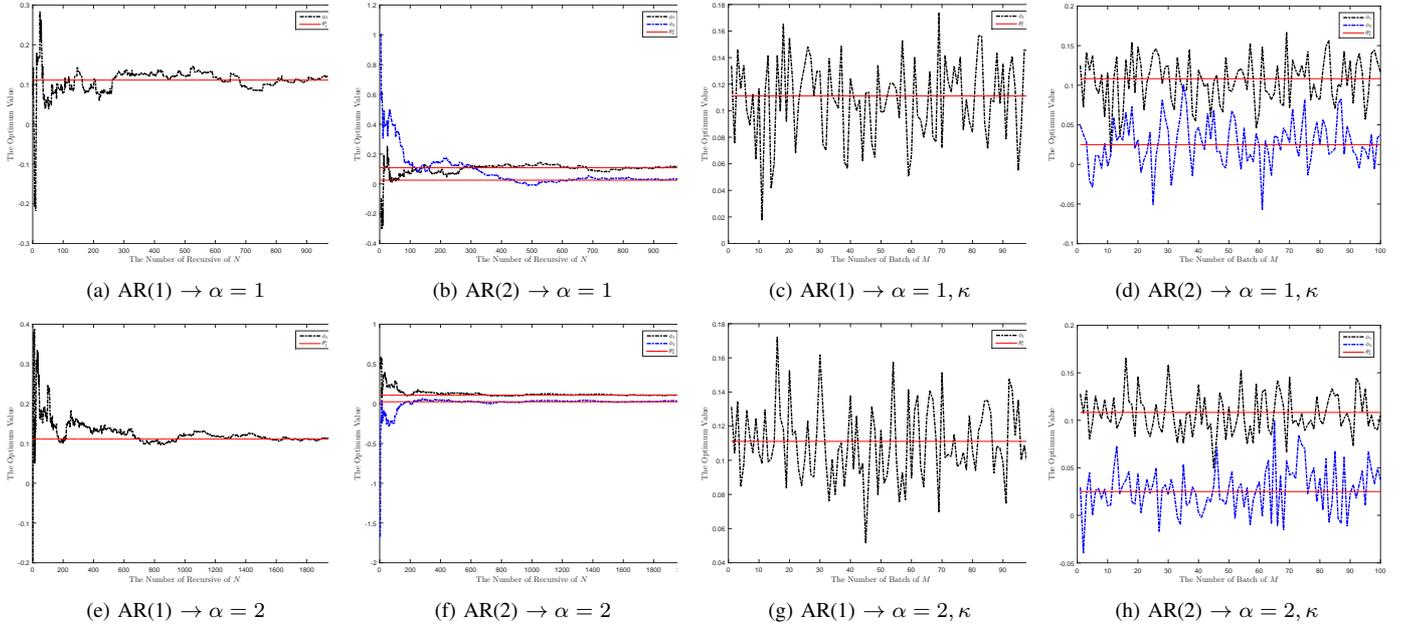

	\centering
	\begin{subfigure}[t]{0.245\linewidth}
		\includegraphics[width=\linewidth]{myplot1.eps}
		\caption{AR(1) $\rightarrow \alpha = 1$}
		\label{fig:a1}
	\end{subfigure}
    \begin{subfigure}[t]{0.245\linewidth}
		\includegraphics[width=\linewidth]{myplot2.eps}
		\caption{AR(2) $\rightarrow \alpha = 1$}
		\label{fig:b1}
	\end{subfigure}
	\begin{subfigure}[t]{0.245\linewidth}
		\includegraphics[width=\linewidth]{myplot3.eps}
		\caption{AR(1) $\rightarrow \alpha = 1, \kappa$}
		\label{fig:c1}    
    \end{subfigure}
    \begin{subfigure}[t]{0.245\linewidth}
		\includegraphics[width=\linewidth]{myplot4.eps}
		\caption{AR(2) $\rightarrow \alpha = 1, \kappa$}
		\label{fig:d1}    
    \end{subfigure}\\
    \medskip
	\begin{subfigure}[t]{0.245\linewidth}
		\includegraphics[width=\linewidth]{myplot5.eps}
		\caption{AR(1) $\rightarrow \alpha = 2$}
		\label{fig:a2}
	\end{subfigure}
    \begin{subfigure}[t]{0.245\linewidth}
		\includegraphics[width=\linewidth]{myplot6.eps}
		\caption{AR(2) $\rightarrow \alpha = 2$}
		\label{fig:b2}
	\end{subfigure}
	\begin{subfigure}[t]{0.245\linewidth}
		\includegraphics[width=\linewidth]{myplot7.eps}
		\caption{AR(1) $\rightarrow \alpha = 2, \kappa$}
		\label{fig:c2}    
    \end{subfigure}
    \begin{subfigure}[t]{0.245\linewidth}
		\includegraphics[width=\linewidth]{myplot8.eps}
		\caption{AR(2) $\rightarrow \alpha = 2, \kappa$}
		\label{fig:d2}    
    \end{subfigure}
    \caption{These show the different performance of either families with parameters of $\alpha$ and $\kappa$}
    \label{Fig:Total1} 
\end{figure*}

\section{Numerical Design}
Linear regression as typical context related to least-square estimator is initialized with $n+1$ variables of $y(t)$, $\theta_1(t), \dots, \theta_n(t)$ over an interval of $t = 1, 2, \dots, N$. It is then required, if possible, $n$ variables $\zeta_1, \zeta_2, \dots, \zeta_n$, such that
\begin{align}
    y(t) = \zeta_1 \theta_1 + \zeta_2 \theta_2 + \dots + \zeta_n \theta_n
\end{align}
is denoted as linear regression according to $\theta(t)$ variable which is able to rewrite as a matrix function of $\zeta^{\top}\theta$ acting as row matrices. Recalling equation (11) with certain iteration of $N$ along with some random sequence of either independent, the recursive equation is set as,
\begin{align}
    \Pi = \begin{bmatrix}
    y(1) & y(2) & \cdots & y(N) \end{bmatrix}
\end{align}
While the appearance of error $\epsilon$ referring to $y(t) - \zeta^{\top}\theta$ is undeniable, to minimize it is a key by obtaining the optimal vector of $\theta^\star$. It can be achieved by the function of quadratic cost, therefore
\begin{align}
    \Gamma(\theta) = \sum_{t = 1}^{N}\left[y(t) - \zeta^{\top}\theta\right]^2 \longrightarrow \theta^{\star} = \arg \min_{\theta} \Gamma(\theta)
\end{align}
such that after differentiating the cost function in terms of $\theta$-element generating the zero results, the $\theta$ could be determined by converting the equality of both row and column vectors. If sum of the product of $\zeta(t)\zeta(t)^\top$ from $1 \to N$ is non-singular,
\begin{align}
    \theta = \left[\sum_{t = 1}^{N}\zeta(t)\zeta(t)^{\top}\right]^{-1} \sum_{t = 1}^{N}\zeta(t)y(t)^{\top}
\end{align}
it can be deduced that for the minimum of $\Gamma$, $\lVert y_N - \zeta_N\theta\rVert$ should be orthogonal to $\zeta_N\theta$. This is due to the symmetric and positive semi-definite so that $\theta_N$ yields in local minimum of $\Gamma(\theta)$. Moreover, the sum of $\zeta(t)\zeta(t)^\top$ results in two divergent scenarios of determinant whether it is zero or not. Unique global minimum or identifiability is situated if it is not zero while for the counterpart result, $\theta_N$ is one among the infinite global minima. The least-square algorithm in Eq. (47) is then used to computed the optimal $\left[\theta_1^{\star},\theta_2^{\star}\right]$ of either families. The variation of $N$ is also applied to observe the effectiveness of the recursive. Furthermore, the $\kappa$ independent batches of the generated $\Pi_i$, such that
\begin{align}
    \Pi_i = \begin{bmatrix}
    y_i(1) & y_i(2) & \dots & y_i(N)
    \end{bmatrix} \quad i = 1\to\kappa
\end{align}
with certain $N$ is required to calculate the empirical mean for both families, such that
\begin{align}
    \bar{\theta}_1 = \frac{1}{\kappa}\sum_{i = 1}^{\kappa}\hat{\theta}_1^{(i)} \qquad \textrm{and} \qquad \bar{\theta}_2 = \frac{1}{\kappa}\sum_{i = 1}^{\kappa}\hat{\theta}_2^{(i)}
\end{align}
along with some variances as follows. The broader batches of $\check{\Pi}_i$ with certain constant $\alpha$, therefore $\alpha N$ is designed.
\begin{align}
    \sigma_{\theta_1}^2\coloneqq\Phi_{\theta_1} = \frac{1}{\kappa}\sum_{i = 1}^{\kappa}\left(\hat{\theta}_1^{(i)} - \bar{\theta}_1\right)^2
\end{align}
\begin{align}
    \sigma_{\theta_2}^2\coloneqq\Phi_{\theta_2} = \frac{1}{\kappa}\sum_{i = 1}^{\kappa}\begin{bmatrix}\left(\varphi_1\right)^2 & \varphi_1\varphi_2\\[7pt]
    \varphi_1\varphi_2 & \left(\varphi_2\right)^2
    \end{bmatrix}
\end{align}
with,
\begin{align}
    \varphi_1 = \hat{\theta}_{2,1}^{(i)} - \bar{\theta}_{2,1} \qquad \varphi_2 = \hat{\theta}_{2,2}^{(i)} - \bar{\theta}_{2,2}
\end{align}
The process $q(\bullet)$ is supposed to be \textit{coloured noise} based on the following recursive equation,
\begin{align}
    q(t) = -\frac{1}{2}q(t-1) + \eta(t)
\end{align}
where $\eta \sim \mathcal{W}\mathcal{N}\left(0, 1\right)$ and $\eta(\bullet)$ is independent from $v(\bullet)$. This scenario is then observed being compared to that of \textit{white noise} in the preceding. It can be concluded that the variance $\Phi_{\theta_2}^{\star}$ is slightly smaller than that of $\Phi_{\theta_1}^{\star}$ meaning that the model approximation to the true system of AR(2) is better than that of AR(1). In terms of the simulation, the yields are closer to the hand-made calculation relative to certain $\alpha N, \alpha = 1\to\infty$, the greater $\alpha$ results in better approximation of optimal value leading to the variance. With $\alpha = [1, 2]$ and $N = 1000$ in AR(1) for instance, it generates
\begin{align}
    0.14 \rightarrow N, \quad 0.12 \rightarrow 2N
\end{align}
optimal value $\theta_1^\star$ respectively compared to the hand calculation $0.11$ while as for AR(2) as $\theta_2^\star$ with the same design of $\alpha$, it yields $[\phi_1, \phi_2]^\top$ as
\begin{align}
    \begin{bmatrix}
    0.128\\
    0.031\end{bmatrix} \rightarrow N, \quad \begin{bmatrix}
    0.123\\
    0.028\end{bmatrix} \rightarrow 2N
\end{align}
in turn against the origin of $[0.108,0.025]^\top$. For certain $\kappa = 100, N = 1000$ and $\alpha = [1,2]$, the $\theta_1^\star$ AR(1) generates,   
\begin{align}
    \begin{bmatrix}
    \theta_1^\star\\
    \Phi_{\theta_1}\end{bmatrix}\coloneqq\begin{bmatrix}
    0.116\\
    0.0011\end{bmatrix}\rightarrow N, \quad \begin{bmatrix}
    0.112\\
    0.0005\end{bmatrix}\rightarrow 2N
\end{align}
and these values of the estimated parameter is close to that of the optimal $\theta_1^\star = 0.11$. This is due to the more data provided leading to better approximation. With respect to AR(2), the optimal parameter $\theta_2^\star$ and the variance $\Phi_{\theta_2}$ with the same scenarios of $\kappa, N, \alpha$ are obtained as follows,
\begin{align}
    \theta_2^\star = \begin{bmatrix}
    0.1048\\
    0.0216\end{bmatrix} \quad \Phi_{\theta_2} = \begin{bmatrix}
    0.00097 & -0.00006\\
    -0.00006 & 0.00111\end{bmatrix}
\end{align}
for $N = 1000$ whereas with $\alpha = 2$, it gains
\begin{align}
    \theta_2^\star = \begin{bmatrix}
    0.1086\\
    0.0245\end{bmatrix} \quad \Phi_{\theta_2} = \begin{bmatrix}
    0.00046 & -0.00002\\
    -0.00002 & 0.00059\end{bmatrix}
\end{align}
compared to the origin of $[\phi_1, \phi_2]^\top = [0.1083, 0.0250]^\top$. Moving to \textit{coloured noise} which generates divergent approximation, for AR(1) with the same design as the \textit{white noise} and hand-mad calculation, the optimal value $\theta_1^\star$ comprises,
\begin{align}
    -0.0137 \rightarrow N, \quad -0.0259 \rightarrow 2N
\end{align}
This not surprising though, as for each sequence iteration, it yields a divergent stimulated data influencing the system so that it is far compared to that of the \textit{white}. As for AR(2), $\theta_2^\star$ is reached as follows,
\begin{align}
    \begin{bmatrix}
    -0.0686\\
    -0.0083\end{bmatrix} \rightarrow N, \quad \begin{bmatrix}
    0.0014\\
    0.0538\end{bmatrix} \rightarrow 2N
\end{align}
which show the same pattern as in AR(1). With respect to batches $\Pi_i\to\kappa$ scenario, the optimal value of AR(1) constitutes beyond the origin meaning that it is biased affecting the empirical and its variances, such that  
\begin{align}
    \begin{bmatrix}
    \theta_1^\star\\
    \Phi_{\theta_1}\end{bmatrix}\coloneqq\begin{bmatrix}
    -0.025\\
    0.0015\end{bmatrix}\rightarrow N, \quad \begin{bmatrix}
    -0.022\\
    0.0005\end{bmatrix}\rightarrow 2N
\end{align}
and AR(2) is also performing the same another convergence number apart from the origin (hand-made calculation), it is proven as in the following,
\begin{align}
    \theta_2^\star = \begin{bmatrix}
    0.1048\\
    0.0216\end{bmatrix} \quad \Phi_{\theta_2} = \begin{bmatrix}
    0.00097 & -0.00006\\
    -0.00006 & 0.00111\end{bmatrix}
\end{align}
for $N = 1000$ whereas with $\alpha = 2$, it gains
\begin{align}
    \theta_2^\star = \begin{bmatrix}
    0.1086\\
    0.0245\end{bmatrix} \quad \Phi_{\theta_2} = \begin{bmatrix}
    0.00046 & -0.00002\\
    -0.00002 & 0.00059\end{bmatrix}
\end{align}

\section{Conclusion}
The mathematical scenarios of switching from zero-mean to non zero-mean along with from white to coloured have been studied to give better picture of the mathematical characteristic from either families. The simulation designs have been proposed in order to observe the effectiveness of certain number of recursive $N$ along with the batches concept $\kappa$ compared to the hand calculation of the optimal values. The greater number of $N$ and batches $\kappa$ with the more complex design of estimate AR(2) results in better approximation of the true system in terms of empirical mean and variance. Further research is to compare another more complex families with modified stochastic scenarios.

\section*{Acknowledgment}
Thanks to Professor Thomas Parisini from the Imperial College London who has taught me in the lecture leading to finishing this paper and to LPDP (Indonesia Endowment Fund for Education) Scholarship from Indonesia.

\bibliographystyle{IEEEtran}
\bibliography{reference.bib}

\begin{thebibliography}{10}
\providecommand{\url}[1]{#1}
\csname url@samestyle\endcsname
\providecommand{\newblock}{\relax}
\providecommand{\bibinfo}[2]{#2}
\providecommand{\BIBentrySTDinterwordspacing}{\spaceskip=0pt\relax}
\providecommand{\BIBentryALTinterwordstretchfactor}{4}
\providecommand{\BIBentryALTinterwordspacing}{\spaceskip=\fontdimen2\font plus
\BIBentryALTinterwordstretchfactor\fontdimen3\font minus
  \fontdimen4\font\relax}
\providecommand{\BIBforeignlanguage}[2]{{%
\expandafter\ifx\csname l@#1\endcsname\relax
\typeout{** WARNING: IEEEtran.bst: No hyphenation pattern has been}%
\typeout{** loaded for the language `#1'. Using the pattern for}%
\typeout{** the default language instead.}%
\else
\language=\csname l@#1\endcsname
\fi
#2}}
\providecommand{\BIBdecl}{\relax}
\BIBdecl

\bibitem{R1}
L.~Ljung, T.~Chen, and B.~Mu, ``A shift in paradigm for system
  identification,'' \emph{International Journal of Control}, pp. 1--8, 2019.

\bibitem{R2}
J.~{Nagumo} and A.~{Noda}, ``A learning method for system identification,''
  \emph{IEEE Transactions on Automatic Control}, vol.~12, no.~3, pp. 282--287,
  June 1967.

\bibitem{R3}
E.~G. {Hurst}, ``Bayesian autoregressive time series analysis,'' \emph{IEEE
  Transactions on Systems Science and Cybernetics}, vol.~4, no.~3, pp.
  317--324, Sep. 1968.

\bibitem{R4}
J.~{Ding}, S.~{Shahrampour}, K.~{Heal}, and V.~{Tarokh}, ``Analysis of
  multistate autoregressive models,'' \emph{IEEE Transactions on Signal
  Processing}, vol.~66, no.~9, pp. 2429--2440, May 2018.

\bibitem{R5}
P.~{Stoica}, ``Prediction of autoregressive lognormal processes,'' \emph{IEEE
  Transactions on Automatic Control}, vol.~25, no.~2, pp. 292--293, April 1980.

\bibitem{R6}
Y.~Nardi and A.~Rinaldo, ``Autoregressive process modeling via the lasso
  procedure,'' \emph{Journal of Multivariate Analysis}, vol. 102, no.~3, pp.
  528 -- 549, 2011.

\bibitem{R7}
R.~K. {Yedavalli}, ``Conditions for the existence of a common quadratic
  lyapunov function via stability analysis of matrix families,'' in
  \emph{Proceedings of the 2002 American Control Conference (IEEE Cat.
  No.CH37301)}, vol.~2, May 2002, pp. 1296--1301 vol.2.

\bibitem{R8}
C.~{Yang}, J.~{Sun}, Q.~{Zhang}, and X.~{Ma}, ``Lyapunov stability and strong
  passivity analysis for nonlinear descriptor systems,'' \emph{IEEE
  Transactions on Circuits and Systems I: Regular Papers}, vol.~60, no.~4, pp.
  1003--1012, April 2013.

\bibitem{R9}
W.~R. Melvin, ``Stability properties of functional difference equations,''
  \emph{Journal of Mathematical Analysis and Applications}, vol.~48, no.~3, pp.
  749 -- 763, 1974.

\bibitem{R10}
M.~V. {Dragosevic} and S.~S. {Stankovic}, ``A generalized least squares method
  for frequency estimation,'' \emph{IEEE Transactions on Acoustics, Speech, and
  Signal Processing}, vol.~37, no.~6, pp. 805--819, June 1989.

\bibitem{R11}
M.~{Morf} and T.~{Kailath}, ``Square-root algorithms for least-squares
  estimation,'' \emph{IEEE Transactions on Automatic Control}, vol.~20, no.~4,
  pp. 487--497, August 1975.

\end{thebibliography}

\newpage

\section{Appendices}
\allowdisplaybreaks
\noindent
\textbf{Appendix A} - $y(t)$
\begin{align*}
	\mathcal{H} =\; & \frac{1}{1 - \lambda z^{-1}}\\
    Y(z) =\; & \mathcal{Q}(z)\mathcal{H}(z) + \mathcal{V}(z)\\
    Y(z) + \frac{Y(z)}{\mathcal{H}(z)} =\; & Y(z) + \mathcal{Q}(z) + \frac{\mathcal{V}(z)}{\mathcal{H}(z)}\\
    Y(z) =\; & Y(z) - \frac{Y(z)}{\mathcal{H}(z)} + \mathcal{Q}(z) + \frac{\mathcal{V}(z)}{\mathcal{H}(z)}\\
    =\; & Y(z) \left[ 1 - \frac{1}{\mathcal{H}(z)} \right] + \mathcal{Q}(z) + \frac{\mathcal{V}(z)}{\mathcal{H}(z)}\\
    =\; & Y(z) \left[\lambda z^{-1} \right] + \mathcal{Q}(z) + \mathcal{V}(z) \left[1 - \lambda z^{-1} \right]\\
    y(t) =\; & \lambda y(t - 1) + q(t) + v(t) - \lambda v(t - 1)
\end{align*}
\noindent
\textbf{Appendix B} - Expected Value $\textbf{E}_y$
\begin{align*}
	\textbf{E}[y(t)] =\; & \textbf{E}[\lambda y(t - 1) + q(t) + v(t) - \lambda v(t - 1)]\\
    =\; & \lambda\textbf{E}[y(t - 1)] + \textbf{E}[q(t)] + \textbf{E}[v(t)] - \lambda\textbf{E}[v(t - 1)]\\
    =\; & \lambda\bar{y} + \bar{q} + \bar{v} - \lambda\bar{v} \\
    =\; & \bar{v} + \dfrac{\bar{q}}{1 - \lambda}
\end{align*}
\noindent
\textbf{Appendix C} - Variance $\Phi_y$
\begin{align*}
	\Phi_y =\; & \textbf{E}\left\lbrace [y(t) - \bar{y}]^2 \right\rbrace\\
    =\; & \textbf{E}\left[y(t)^2 - 2y(t)\bar{y} + \bar{y}^2\right] \\
    =\; & \textbf{E}\left[y(t)^2\right] - \textbf{E}\left[\bar{y}^2\right] \longrightarrow \textbf{E}\left[\bar{y}^2\right] = 0\\
    =\; & \textbf{E}\left\lbrace \left[\underbrace{\lambda y(t - 1) + q(t)}_{A} + \overbrace{v(t) - \lambda v(t - 1)}^{B} \right]^2 \right\rbrace\\
    =\; & \textbf{E}\left[A^2 + 2AB + B^2\right]\\
    \textbf{E}\left[A^2\right] =\; & \textbf{E}\left[\lambda^2 y(t - 1)^2 + 2\lambda y(t - 1)q(t) + q(t)^2\right]\\
    =\; & \textbf{E}\left[\lambda^2 y(t - 1)^2\right] + 2\textbf{E}\left[\lambda y(t - 1)q(t)\right] + \textbf{E}\left[q(t)^2\right]\\
    =\; & \lambda^2 \Phi_y + \delta^2\\
    \textbf{E}\left[2AB\right] =\; & 2\textbf{E}\left[\lambda y(t - 1)v(t) - \lambda^2 y(t - 1)v(t - 1) + q(t)v(t)\right. \\
    \; & -\lambda q(t)v(t - 1)]\\
    =\; & 2\textbf{E}\left[\lambda y(t - 1)v(t)\right] - 2\lambda^2 \textbf{E}\left[y(t - 1)v(t - 1)\right] \\
    \; & +2\textbf{E}\left[q(t)v(t)\right] - 2\textbf{E}\left[\lambda q(t)v(t - 1)\right]\\
    =\; & -2 \lambda^2 \xi^2\\
    \textbf{E}\left[B^2\right] =\; & \textbf{E}\left[v(t)^2 - 2\lambda v(t)v(t - 1) + \lambda^2v(t - 1)^2\right]\\
    =\; & \textbf{E}\left[v(t)^2\right] - 2\lambda\textbf{E}\left[v(t)v(t - 1)\right] + \lambda^2\textbf{E}\left[v(t - 1)^2\right]\\
    =\; & \xi^2 + \lambda^2 \xi^2\\
    \Phi_y =\; & \lambda^2 \Phi_y + \delta^2 - 2\lambda^2 \xi^2 + \xi^2 + \lambda^2 \xi^2\\
    =\; & \frac{\delta^2}{1 - \lambda^2} + \xi^2
\end{align*}
\noindent
\textbf{Appendix D} - Covariance $\Psi_y$
\begin{align*}
	\Psi_y(\tau) =\; & \textbf{E}\Bigl\lbrace\left[y(t) - \bar{y}\right]\left[y(t - \tau) - \bar{y}\right]\Bigl\rbrace \longrightarrow \bar{y} = 0\\
    =\; & \textbf{E}\Bigl\lbrace\left[ay(t - 1) + e(t) + v(t) - av(t - 1)\right]\left[y(t - \tau)\right]\Bigl\rbrace\\
    =\; & \lambda\textbf{E}\left[y(t - 1)y(t - \tau)\right] + \textbf{E}\left[e(t)y(t - \tau)\right] \\
    \; & +\textbf{E}\left[v(t)y(t - \tau)\right] - \lambda\textbf{E}\left[v(t - 1)y(t - \tau)\right]\\
    \Psi_y(0) =\; & \lambda\Psi(1) + \delta^2 + \xi^2\\
    \Psi_y(1) =\; & \lambda\Psi(0) - \lambda\xi^2\\
    \Psi_y(2) =\; & \lambda\Psi(1)
\end{align*}
\noindent
\textbf{Appendix D$^\star$} - Covariance $\Psi_y(\tau)\longrightarrow \tau = 0,1,2$
\begin{align*}
    \Psi_y(1) \coloneqq\; & \frac{\Psi(1)}{\lambda} + \xi^2 = \lambda\Psi(1) + \delta^2 + \xi^2\\
    \coloneqq\; & \frac{\lambda\delta^2}{1 - \lambda^2}\\
    \Psi_y(0) \coloneqq\; & \lambda\frac{\lambda\delta^2}{1 - \lambda^2} + \delta^2 + \xi^2\\
    \coloneqq\; & \frac{\delta^2}{1 - \lambda^2} + \xi^2 \longrightarrow \Phi_y\\
    \Psi_y(2) \coloneqq\; & \lambda\Psi(1)\\
    \coloneqq\; & \frac{\lambda^2\delta^2}{1 - \lambda^2} \longrightarrow \frac{\lambda^{\tau}\delta^2}{1 - \lambda^2} \textrm{ for } \Psi_y(\tau > 1)
\end{align*}
\noindent
\textbf{Appendix E} - $\Gamma(\theta_1) \longrightarrow \phi_1$
\begin{align*}
    \Gamma(\theta_1) =\; & \textbf{E}\left\lbrace\left[y(t) - \hat{y}(t|t-1)\right]^2\right\rbrace\\
    =\; & \textbf{E}\left\lbrace\left[y(t) - \phi_1 y(t-1)\right]^2\right\rbrace\\
    =\; & \textbf{E}\left[y(t)^2 - 2\phi_1 y(t)y(t-1) + \phi^2_1 y(t-1)^2\right]\\
    =\; & \Psi_y(0) - 2\phi_1\Psi_y(1) + \phi^2_1\Psi_y(0)\\
    =\; & \left(1 + \phi^2_1\right)\Psi_y(0) - 2\phi_1\Psi_y(1) \\
    \frac{\partial\Gamma(\theta_1)}{\partial\phi_1} =\; & 2\phi_1\Psi_y(0) - 2\Psi_y(1) = 0\\
    \phi_1 =\; & \frac{\Psi_y(1)}{\Psi_y(0)} \longrightarrow \frac{\lambda\Psi_y(0) - \lambda\xi^2}{\Psi_y(0)}\\
    =\; & \frac{\lambda\delta^2}{\delta^2 + (1-\lambda^2)\xi^2}
\end{align*}
\noindent
\textbf{Appendix F} - $\Gamma(\theta_2) \longrightarrow \theta_2^{\star} \coloneqq \left[\phi_1, \phi_2\right]^{\top}$
\begin{align*}
    \Gamma(\theta_2) =\; & \textbf{E}\left\lbrace\left[y(t) - \hat{y}(t|t-1)\right]^2\right\rbrace\\
    =\; & \textbf{E}\left\lbrace\left[y(t) - \phi_1 y(t-1) - \phi_2 y(t-2)\right]^2\right\rbrace\\
    =\; & \textbf{E}\left[y(t)^2 + \phi^2_1 y(t-1)^2 + \phi^2_2 y(t-2)^2 \right. \\
    \; & -2\phi_1 y(t)y(t-1) - 2\phi_2 y(t)y(t-2) \\
    \; & +2\phi_1\phi_2 y(t-1)y(t-2)]\\
    =\; & \Psi_y(0) + \phi_1^2\Psi_y(0) + \phi_2^2\Psi_y(0) - 2\phi_1\Psi_y(1) \\
    \; & -2\phi_2\Psi_y(2) + 2\phi_1\phi_2\Psi_y(1)\\
    =\; & \left(1 + \phi^2_1 + \phi^2_2\right)\Psi_y(0) + 2\phi_1\left(\phi_2 - 1\right)\Psi_y(1) \\
    \; &- 2\phi_2\Psi_y(2)\\
    \frac{\partial\Gamma(\theta_2)}{\partial\phi_1} =\; & 2\phi_1\Psi_y(0) + 2 \phi_2\Psi_y(1) - 2\Psi_y(1) = 0\\
    \phi_1 =\; & \frac{\Psi_y(1) - \phi_2\Psi_y(1)}{\Psi_y(0)} \\
    =\; & \frac{\Psi_y(1) - \left(\dfrac{\Psi_y(2)\Psi_y(0) - \Psi_y(1)^2}{\Psi_y(0)^2 - \Psi_y(1)^2}\right)\Psi_y(1)}{\Psi_y(0)}\\
    =\; & \frac{\Psi_y(1)\Psi_y(0) - \Psi_y(2)\Psi_y(1)}{\Psi_y(0)^2 - \Psi_y(1)^2}\\
    =\; & \frac{\lambda\delta^2\left(\delta^2 + \xi^2\right)}{\left(\delta^2 + \xi^2\right)^2 - \lambda^2\xi^4}\\
    \frac{\partial\Gamma(\theta_2)}{\partial\phi_1} =\; & 2\phi_2\Psi_y(0) + 2\phi_1\Psi_y(1) - 2\Psi_y(2) = 0\\
    \phi_2 =\; & \frac{\Psi_y(2) - \phi_1\Psi_y(1)}{\Psi_y(0)} \\
    =\; & \frac{\Psi_y(2) - \left(\dfrac{\Psi_y(1)\Psi_y(0) - \Psi_y(2)\Psi_y(1)}{\Psi_y(0)^2 - \Psi_y(1)^2}\right)\Psi_y(1)}{\Psi_y(0)} \\
    =\; & \frac{\Psi_y(2)\Psi_y(0) - \Psi_y(1)^2}{\Psi_y(0)^2 - \Psi_y(1)^2}\\
    =\; & \frac{\lambda^2\delta^2\xi^2}{\left(\delta^2 + \xi^2\right)^2 - \lambda^2\xi^4}
\end{align*}
\noindent
\textbf{Appendix G} - $\Phi_{\theta_1^{\star}}$ and $\Phi_{\theta_2^{\star}}$
\begin{align*}
    \Phi_{\theta_1^{\star}} =\; & \Gamma(\theta_1) = \textbf{E}\left\lbrace\left[\varepsilon_{\theta_1}(t)-\bar{\varepsilon}_{\theta_1}\right]^2\right\rbrace \longrightarrow \bar{\varepsilon}_{\theta_1} = 0\\
    =\; & \left[1 + \phi_1^2\right]\Psi_y(0) - 2\phi_1\Psi_y(1) \longrightarrow \phi_1 \textrm{ from Eq. (21)}\\
    =\; & \left[1 + \left(\frac{\lambda\delta^2}{\delta^2 + (1-\lambda^2)\xi^2}\right)^{2}\right]\frac{\delta^2 + (1 - \lambda^2)\xi^2}{1 - \lambda^2} \\
    \; & -2\left(\frac{\lambda\delta^2}{\delta^2 + (1-\lambda^2)\xi^2}\right)\frac{\lambda\delta^2}{1 - \lambda^2}\\
    =\; & \frac{\left[\delta^2 + \left(1 - \lambda^2\right)\xi^2\right]^2 - \lambda^2\delta^4}{\left[\delta^2 + \left(1-\lambda^2\right)\xi^2\right]\left[1-\lambda^2\right]}\\
    =\; & \frac{\delta^4\left(1-\lambda^2\right) + 2\delta^2\xi^2 \left(1-\lambda^2\right) + \xi^4\left(1-\lambda^2\right)^2}{\left[\delta^2 + \left(1-\lambda^2\right)\xi^2\right]\left[1-\lambda^2\right]}\\
    =\; & \frac{\left(\delta^2 + \xi^2\right)^2 - \lambda^2\xi^4}{\delta^2 + \left(1-\lambda^2\right)\xi^2}\\
    \Phi_{\theta_2^{\star}} =\; & \Gamma(\theta_2) = \textbf{E}\left\lbrace\left[\varepsilon_{\theta_2}(t)-\bar{\varepsilon}_{\theta_2}\right]^2\right\rbrace \longrightarrow \bar{\varepsilon}_{\theta_2} = 0\\
    =\; & \left[1 + \phi_1^2 + \phi_2^2\right]\Psi_y(0) - \left[2\phi_1 - 2\phi_1\phi_2\right] \Psi_y(1) \\
    \; & -2\phi_2\Psi_y(2) \longrightarrow [\phi_1, \phi_2] \textrm{ from Eq. (22)}\\
    =\; & \Psi_y(0)^3 - \Psi_y(2)^2\Psi_y(0) - 2\Psi_y(1)^2\Psi_y(0) \\
    \; & +2\Psi_y(1)^2\Psi_y(2) \longrightarrow \textrm{after some complex-algebra}\\
    =\; & \left(\frac{\delta^2 + (1 - \lambda^2)\xi^2}{1 - \lambda^2}\right)^3 - \left(\frac{\lambda^2\delta^2}{1 - \lambda^2}\right)^2\frac{\delta^2 + (1 - \lambda^2)\xi^2}{1 - \lambda^2} \\
    \; & -2\left(\frac{\lambda\delta^2}{1 - \lambda^2}\right)^2\frac{\delta^2 + (1 - \lambda^2)\xi^2}{1 - \lambda^2} + 2\lambda\left(\frac{\lambda\delta^2}{1 - \lambda^2}\right)^3\\
    =\; & \frac{\left(\delta^2 + \xi^2\right)^3 - \lambda^2 \left(\delta^2\xi + \xi^3\right)^2 + 2\lambda\delta^4\xi^2}{1 - \lambda^2}
\end{align*}
\noindent
\textbf{Appendix H} - $\Gamma(\theta_1) \longrightarrow \phi_1$; (non) zero-mean
\begin{align*}
    \Gamma(\theta_1) =\; & \textbf{E}\left[y(t)^2 - 2y(t)y(t|t-1) +  y(t|t-1)^2\right]\\
    =\; & \textbf{E}\left\lbrace\left[\check{y}(t)+\bar{y}\right]^2\right\rbrace - 2\phi_1\textbf{E}\Bigl\lbrace\left[\check{y}(t)+\bar{y}\right] \left[\check{y}(t-1)+\bar{y}\right]\Bigl\rbrace\\
    \; & + \phi_1^2\textbf{E}\left\lbrace\left[\check{y}(t-1)+\bar{y}\right]^2\right\rbrace\\
    =\; & \left[1 + \phi_1^2\right]\Psi_y(0) - 2\phi_1\Psi_y(1) + \left(1-\phi_1\right)^2\bar{y}^2\\
    \frac{\partial\Gamma(\theta_1)}{\partial\phi_1} =\; & -2\Psi_y(1) + 2\phi_1\Psi_y(0) - 2\left(1-\phi_1\right)\bar{y}^2 = 0\\
    \theta^{\star}_1 \coloneqq\; & \frac{\Psi(1) + \bar{y}^2}{\Psi(0) + \bar{y}^2}
\end{align*}
\noindent
\textbf{Appendix I} - $\Gamma(\theta_2) \longrightarrow \theta_2^{\star} \coloneqq \left[\phi_1, \phi_2\right]^{\top}$; (non) zero-mean
\begin{align*}
    \Gamma(\theta_2) =\; & \textbf{E}\left[y(t)^2 - 2y(t)y(t|t-1) +  y(t|t-1)^2\right]\\
    =\; & \textbf{E}\left\lbrace\left[\check{y}(t)+\bar{y}\right]^2\right\rbrace - 2\textbf{E}\Bigl\lbrace\left[\check{y}(t)+\bar{y}\right] \left[\phi_1\check{y}(t-1)\right.\Bigl.\\
    \; & +\Bigl.\left.\phi_2\check{y}(t-2)+\left(\phi_1+\phi_2\right)\bar{y}\right]\Bigl\rbrace + \phi_1^2\textbf{E}\left\lbrace\left[\check{y}(t-1)\right]^2\right\rbrace\\
    \; & + \phi_2^2\textbf{E}\left\lbrace\left[\check{y}(t-2)\right]^2\right\rbrace + \left(\phi_1+\phi_2\right)\textbf{E}\left\lbrace\left[\bar{y}\right]^2\right\rbrace\\
    \; & + 2\phi_1\left(\phi_1+\phi_2\right)\textbf{E}\left\lbrace\left[\check{y}(t-1)\bar{y}\right]\right\rbrace\\
    \; & + 2\phi_2\left(\phi_1+\phi_2\right)\textbf{E}\left\lbrace\left[\check{y}(t-2)\bar{y}\right]\right\rbrace\\
    \; & + 2\phi_1\phi_2\textbf{E}\left\lbrace\left[\check{y}(t-1)\check{y}(t-2)\right]\right\rbrace\\
    =\; & \Psi_y(0) + \phi_1^2\Psi_y(0) + \phi_2^2\Psi_y(0) - 2\phi_1\Psi_y(1) \\
    \; & -2\phi_2\Psi_y(2) + 2\phi_1\phi_2\Psi_y(1) + \left[1-\left(\phi_1+\phi_2\right)\right]^2\bar{y}^2\\
    \frac{\partial\Gamma(\theta_2)}{\partial\phi_1} =\; & 0 \textrm{ and } \frac{\partial\Gamma(\theta_2)}{\partial\phi_2} =\; 0
\end{align*}
\textrm{after some complex-algebra}
\begin{align*}
    \phi_1 =\; & \dfrac{\Psi_y(1)\Psi_y(0) - \Psi_y(2)\Psi_y(1) + \bar{y}^2 \left[\Psi(0) - \Psi(2)\right]}{\Psi_y(0)^2 - \Psi_y(1)^2 + 2\bar{y}^2 \left[\Psi(0) - \Psi(1)\right]} \\
    \phi_2 =\; & \dfrac{\Psi_y(2)\Psi_y(0) - \Psi_y(1)^2 + \bar{y}^2 \left[\Psi(0) - 2\Psi(1) + \Psi(2)\right]}{\Psi_y(0)^2 - \Psi_y(1)^2 + 2\bar{y}^2 \left[\Psi(0) - \Psi(1)\right]}
\end{align*}
\noindent
\textbf{Appendix J} - $\Phi_{\theta_1^{\star}}$ and $\Phi_{\theta_2^{\star}}$; (non) zero-mean
\begin{align*}
    \bar{\varepsilon}_{\theta_1} =\; & \textbf{E}\left[\check{y}(t)\right] + \textbf{E}\left[\bar{y}\right] - \phi_1\textbf{E}\left[\check{y}(t-1)\right] - \phi_1\textbf{E}\left[\bar{y}\right]\\
    =\; & (1 - \phi_1)\bar{y}\\
    \Phi_{\theta_1^{\star}} =\; & \Gamma(\theta_1) = \textbf{E}\left\lbrace\left[\varepsilon_{\theta_1}(t)-\bar{\varepsilon}_{\theta_1}\right]^2\right\rbrace\\
    =\; & \textbf{E}\left\lbrace\left[\check{y}(t) + \bar{y} - \phi_1 \check{y}(t-1) - \phi_1\bar{y} - \left(1-\phi_1\right)\bar{y} \right]^2\right\rbrace\\
    =\; & \textbf{E}\left\lbrace\left[\check{y}(t) - \phi_1 \check{y}(t-1)\right]^2\right\rbrace \longrightarrow \textrm{\textbf{Appendix E}}\\
    =\; & \frac{\left(\delta^2 + \xi^2\right)^2 - \lambda^2\xi^4}{\delta^2 + \left(1-\lambda^2\right)\xi^2} \longrightarrow \textrm{\textbf{Appendix G}}\\
    \bar{\varepsilon}_{\theta_2} =\; & \textbf{E}\left[\check{y}(t)\right] + \textbf{E}\left[\bar{y}\right] - \phi_1\textbf{E}\left[\check{y}(t-1)\right] - \phi_2\textbf{E}\left[\check{y}(t-2)\right]\\
    \; & - \left(\phi_1 + \phi_2\right) \textbf{E}\left[\bar{y}\right]\\
    =\; & (1 - \phi_1 - \phi_2)\bar{y}\\
    \Phi_{\theta_2^{\star}} =\; & \Gamma(\theta_2) = \textbf{E}\left\lbrace\left[\varepsilon_{\theta_2}(t)-\bar{\varepsilon}_{\theta_2}\right]^2\right\rbrace\\
    =\; & \textbf{E}\Bigl\lbrace\left[\check{y}(t) + \bar{y} - \phi_1 \check{y}(t-1) - \phi_2 \check{y}(t-2) \right.\Bigl.\\
    \; & \Bigl.\left.- \left(\phi_1 + \phi_2\right)\bar{y} - \left(1-\phi_1-\phi_2\right)\bar{y} \right]^2\Bigl\rbrace\\
    =\; & \textbf{E}\left\lbrace\left[\check{y}(t) - \phi_1 \check{y}(t-1) - \phi_2 \check{y}(t-2)\right]^2\right\rbrace\\
    =\; & \frac{\left(\delta^2 + \xi^2\right)^3 - \lambda^2 \left(\delta^2\xi + \xi^3\right)^2 + 2\lambda\delta^4\xi^2}{1 - \lambda^2}
\end{align*}

\section*{Authors}
\noindent\textbf{First and Correspondence Author} - Moh Kamalul Wafi was graduated from the Imperial College London majoring Control Systems under Department of Electrical and Electronics Engineering. I am currently with the Laboratory of Embedded and Cyber-Physical Systems, Department of Engineering Physics, Institut Teknologi Sepuluh Nopember, 60111, Indonesia

\end{document}